\documentclass[letterpaper, 10 pt, conference]{ieeeconf}  

\IEEEoverridecommandlockouts                              
\usepackage{amsmath} 
\usepackage{amssymb}  
\usepackage{graphicx,cite}
\usepackage{mathrsfs}
\usepackage{algorithm,algorithmic}
\usepackage{url}

\usepackage[caption=false,font=footnotesize]{subfig}

\newcommand{\Real}[1]{{\rm Re}\,(#1)}
\newcommand{\Imag}[1]{{\rm Im}\,(#1)}
\newcommand{\trace}{{\rm tr}}

\usepackage{theorem}
\theorembodyfont{\rmfamily}

\renewcommand{\QED}{\QEDopen}

\title{\LARGE \bf
Adversarial Attacks to Direct Data-driven Control for Destabilization
}

\author{Hampei Sasahara
\thanks{This work was supported by JSPS KAKENHI Grant Number 22K21272.}
\thanks{H. Sasahara is with the Department of Systems and Control Engineering, School of Engineering, Tokyo Institute of Technology, Tokyo 152-8552, Japan
        {\tt\small sasahara@sc.e.titech.ac.jp}}
\thanks{The MATLAB codes are available at \protect\url{http://hampei.net/codes/HS23_MATLAB_AdvAttacks.zip}}
}

\begin{document}

\maketitle
\thispagestyle{empty}
\pagestyle{empty}

\begin{abstract}
This study investigates the vulnerability of direct data-driven control to adversarial attacks in the form of a small but sophisticated perturbation added to the original data.
The directed gradient sign method (DGSM) is developed as a specific attack method, based on the fast gradient sign method (FGSM), which has originally been considered in image classification.
DGSM uses the gradient of the eigenvalues of the resulting closed-loop system and crafts a perturbation in the direction where the system becomes less stable.
It is demonstrated that the system can be destabilized by the attack, even if the original closed-loop system with the clean data has a large margin of stability.
To increase the robustness against the attack, regularization methods that have been developed to deal with random disturbances are considered.
Their effectiveness is evaluated by numerical experiments using an inverted pendulum model.
\end{abstract}

\section{INTRODUCTION}

Advances in computing power and an increase in available data have led to the success of data-driven methods in various applications, such as autonomous driving~\cite{Kiran2022Deep}, communications~\cite{Luong2019Applications}, and games~\cite{Silver2016Mastering}.
In the field of control theory, this success has sparked a trend towards direct data-driven control methods, which aim at designing controllers directly from data without the need for a system identification process~\cite{Campi2006Direct,Lewis2013Reinforcement,Mohammadi2021Convergence}.
One prominent scheme is the Willems' fundamental lemma-based approach~\cite{De2019Formulas}, which provides explicit control formulations and requires low computational complexity~\cite{De2021Low}.

Meanwhile, in the context of image classification, it has been reported that a data-driven method using neural networks is susceptible to adversarial attacks~\cite{Bruna2014Intriguing,Goodfellow2015Explaining,Akhtar2018Threat}.
Specifically, adding small perturbations to images that remain imperceptible to human vision system can change the prediction of the trained neural network classifier.
This type of vulnerability has also been observed in different domains such as speech recognition~\cite{Carlini2016Hidden} and reinforcement learning~\cite{Huang2017Adversarial}.
Influenced by those results, adversarial attacks and defenses have become a critical area of research on data-driven techniques.

Most work on control system security focuses on vulnerabilities of control systems themselves and defense techniques with explicit model knowledge against attacks exploiting these vulnerabilities, such as zero-dynamics attack analysis~\cite{Pasqualetti2015Control}, observer-based attack detection~\cite{Giraldo2018A}, and moving target defense~\cite{Griffioen2020Moving}.
In addition, there have also been recent studies on data-driven approaches, such as data-driven stealthy attack design~\cite{Alisic2021Data,Alisic2023Model} and data-driven attack detection~\cite{Krishnan2020Data}.
However, the vulnerability of data-driven control algorithm has received less attention, and there is a need for dedicated techniques to address this issue.

The main objective of this study is to evaluate the robustness of direct data-driven control methods against adversarial attacks, and to provide insights on how to design secure and reliable data-driven controller design algorithms.
The aim of the attacker is to disrupt the stability of the closed-loop system by making small modifications to the data.
As the worst-case scenario, we first consider a powerful attacker who has complete knowledge of the system, the controller design algorithm, and the clean input and output data.
Subsequently, we consider gray box attacks where we assume that the adversary has access to the model and the algorithm but not the data, and additionally may not know design parameters in the algorithm.
Effectiveness of crafted perturbations without partial knowledge is known as \emph{the transferability property,} which has been confirmed in the domain of computer vision~\cite{Szegedy2014Intriguing} and reinforcement learning~\cite{Huang2017Adversarial}.
We observe that the data and parameter transferability property holds in direct data-driven control as well.

\begin{figure*}[t]
  \centering
  \includegraphics[width=0.98\linewidth]{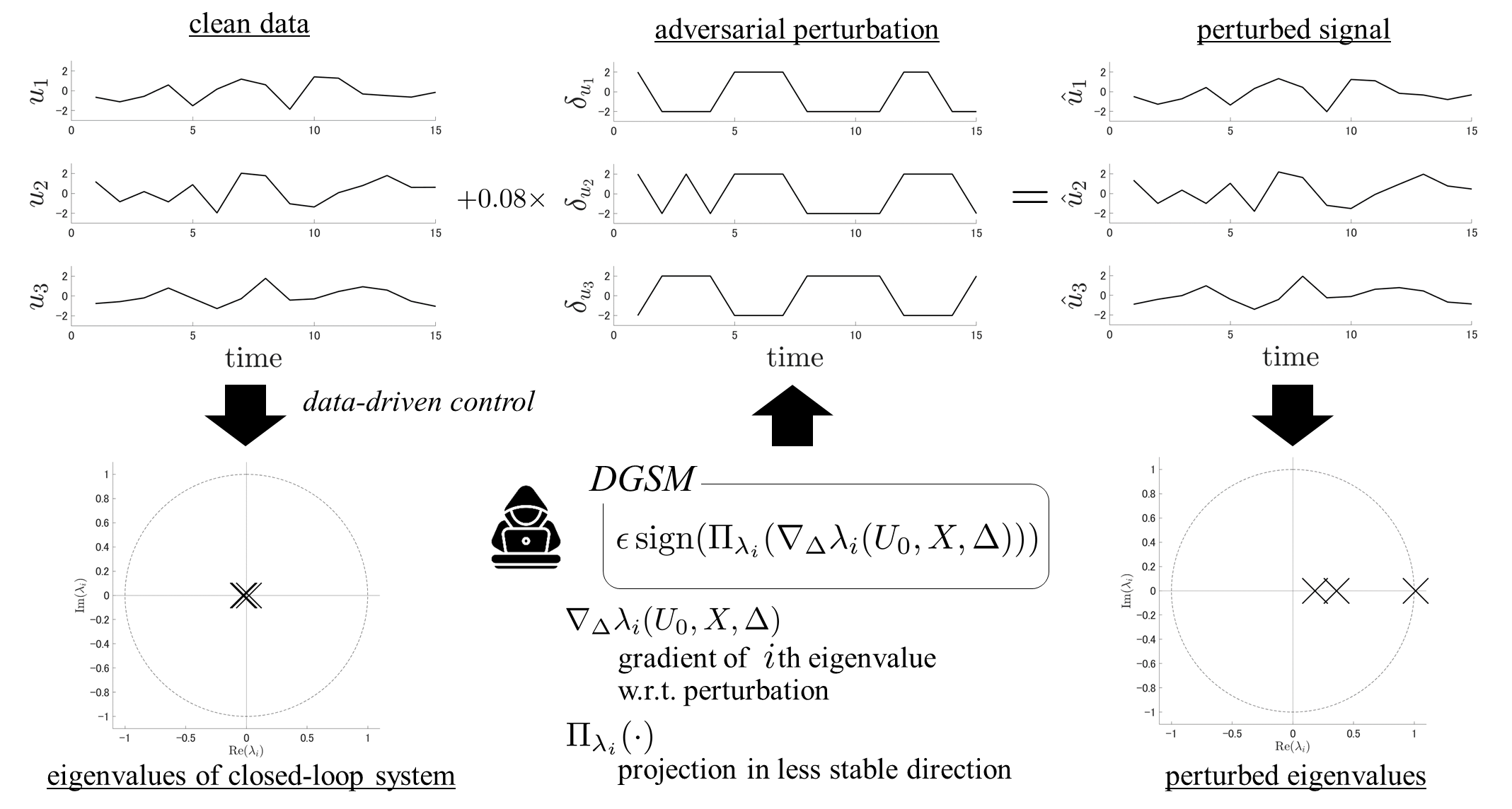}
  \caption{
  Demonstration of DGSM applied to a discrete-time linear system with three-dimensional input.
  The adversarial perturbation created by DGSM is added to the original signal, but the perturbed signal appears almost identical to the original one.
  Nevertheless, the resulting closed-loop system obtained through direct data-driven control with a regularizer becomes unstable due to the adversarial attack.
  Indeed, the eigenvalues of the closed-loop system with the clean data are $\{-0.0177, 0.0212, -0.0275\}$, while those with the perturbed data are $\{0.1824, 0.3613, 1.0120\}$.
  The specific parameters of this instance are provided in Appendix.
  Note that the output signal is also perturbed but its illustration is omitted for clarity.
  }
  \label{fig:vuln_ins}
\end{figure*}

Our first contribution is to demonstrate the vulnerability of direct data-driven control.
We introduce a specific attack, which we refer to as the directed gradient sign method (DGSM), based on the fast gradient sign method (FGSM), which has originally been developed for efficient computation of a severe adversarial perturbation in image classification~\cite{Goodfellow2015Explaining}.
The idea behind FGSM is to calculate the perturbation vector in the direction of the gradient of the cost function while limiting each element's absolute value to a specified small constant.
DGSM is an adaptation of this method, designed to destabilize the targeted control system.
DGSM calculates the gradient of the eigenvalues of the resulting closed-loop system and determines the perturbation in the direction that makes the system less stable.
Fig.~\ref{fig:vuln_ins} illustrates a demonstration of DGSM applied to a discrete-time linear system.
It is shown that while the system can be stabilized by using clean data where the resulting eigenvalues are far from the unit circle it can be made unstable by a small but sophisticated perturbation.

Second, we investigate defense methods using regularization.
We consider two regularization approaches: the first is the certainty-equivalence regularization that links the direct data-driven control with the indirect one via system identification using the ordinary least-square estimation~\cite{Dorfler2022On,Dorfler2022Bridging}.
The second is the robustness-inducing regularization that ensures robustness against noise~\cite{De2021Low}.
We demonstrate that both approaches can improve robustness against adversarial attacks and compare their effectiveness.


\subsection*{Organization and Notation}
The paper is organized as follows.
Sec.~\ref{sec:pre} reviews key concepts of direct data-driven control based on the fundamental lemma and discusses a technique for generating adversarial perturbations used in image classification with neural networks.
In Sec.~\ref{sec:attack}, we outline the attack scenario and present the adversarial method adapted for direct data-driven control that leads to destabilization.
Sec.~\ref{sec:ex} provides experimental evaluation to discuss the vulnerabilities of interest and the improvement in robustness through regularization.
Finally, Sec.~\ref{sec:conc} concludes and summarizes the paper.

We denote the transpose of a matrix $M$ by $M^{\sf T}$,
the trace and the spectrum of a square matrix $M$ by $\trace(M)$ and $\sigma(M)$, respectively,
the maximum and minimum singular values of a matrix $M$ by $\sigma_{\rm max}(M)$ and $\sigma_{\rm min}(M),$ respectively,
the max norm of a matrix $M$ by $\|M\|_{\rm max}$,
the right inverse of a right-invertible matrix $M$ by $M^{\dagger}$,
the positive and negative (semi)definiteness of a Hermetian matrix $M$ by $M\succ $ ($\succeq$) $0$ and $M\prec$ ($\preceq$) $0$, respectively,
and the component-wise sign function by ${\rm sign}(\cdot)$.

\section{PRELIMINARY}
\label{sec:pre}

\subsection{Data-Driven Control based on Fundamental Lemma}
\label{subsec:data-driven}
We first review the direct data-driven control based on the Willems' fundamental lemma~\cite{De2019Formulas}.
Consider a discrete-time linear time-invariant system
$x(t+1)=Ax(t)+Bu(t)+d(t)$ for $t\in\mathbb{N}$
where $x(t)\in\mathbb{R}^n$ is the state, $u(t)\in\mathbb{R}^m$ is the control input, and $d(t)\in\mathbb{R}^n$ is the exogenous disturbance.
Assume that the pair $(A,B)$ is unknown to the controller designer but it is stabilizable.
We consider the linear quadratic regulator (LQR) problem~\cite[Chap.~6]{Chen2012Optimal}, which has widely been studied as a benchmark problem.
Specifically, design a static state-feedback control $u(t)=Kx(t)$ that minimizes the cost function $J(K)=\sum_{i=1}^n\sum_{t=0}^{\infty} \{x(t)^{\sf T}Qx(t) + u(t)^{\sf T}Ru(t)\}|_{x(0)=e_i}$
with $Q\succeq 0$ and $R\succ 0$ where $e_i$ is the $i$th canonical basis vector.
It is known that the cost function can be rewritten as
$J(K)=\trace(QP)+\trace(K^{\sf T}RKP)$
where $P\succeq I$ is the controllability Gramian of the closed-loop system when $A+BK$ is Schur.

The objective of direct data-driven control is to design the optimal feedback gain using data of input and output signals without explicit system identification.
Assume that the time series
$ U_0 :=[u(0)\ u(1)\ \cdots u(T-1)]\in \mathbb{R}^{m\times T}$
and
$X :=[x(0)\ x(1)\ \cdots x(T-1)\ x(T)] \in \mathbb{R}^{n\times (T+1)}$
are available.
The first and last $T$-long time series of $X$ are denoted by $X_0\in \mathbb{R}^{m\times T}$ and $X_1\in \mathbb{R}^{m\times T},$ respectively.
Letting $D_0:=[d(0)\ d(1)\ \cdots\ d(T-1)]\in \mathbb{R}^{m\times T},$
we have the relationship
\[
 X_1-D_0 = [B\ A]W_0,
\]
where
$W_0:=[U_0^{\sf T}\ X_0^{\sf T}]^{\sf T}$.
We here assume that ${\rm rank}\, W_0 = n+m$
holds.
This rank condition, which is generally necessary for data-driven LQR design~\cite{Van2020Data}, is satisfied if the input signal is persistently exciting in the noiseless case as shown by the Willems' fundamental lemma~\cite{Willems2005Note}.

The key idea of the approach laid out in~\cite{De2019Formulas} is to parameterize the controller using the available data by introducing a new variable $G\in\mathbb{R}^{T\times n}$ with the relationship
\begin{equation}\label{eq:KI}
[K^{\sf T}\ I]^{\sf T}
=W_0G.
\end{equation}
Then the closed-loop matrix can be parameterized directly by data matrices as
$A+BK = [B\ A]W_0G = (X_1-D_0)G.$
The LQR controller design can be formulated as
\begin{equation}\label{eq:prob_ori}
\begin{array}{cl}
 \displaystyle{\min_{P,K,G}} & \trace(QP)+\trace(K^{\sf T}RKP)\\
 {\rm s.t.} & X_1GPG^{\sf T}X_1^{\sf T}-P+I\preceq 0\\
  & P \succeq I\ {\rm and}\ \eqref{eq:KI}\\
  & 
\end{array}
\end{equation}
by disregarding the noise term.

However, it has been revealed that the formulation~\eqref{eq:prob_ori} is not robust to disturbance~\cite{Dorfler2022On}.
To enhance robustness against disturbance, a regularized formulation has been proposed:
\begin{equation}\label{eq:prob_reg1}
\begin{array}{cl}
 \displaystyle{\min_{P,K,G}} & \trace(QP)+\trace(K^{\sf T}RKP)+\gamma\|\Pi G\|\\
 {\rm s.t.} & X_1GPG^{\sf T}X_1^{\sf T}-P+I\preceq 0\\
  & P \succeq I\ {\rm and}\ \eqref{eq:KI}\\
\end{array}
\end{equation}
with a constant $\gamma\geq0$ where $\Pi:=I-W_0^{\dagger}W_0$
and $\|\cdot\|$ is any matrix norm.
The regularizer $\gamma\|\Pi G\|$ is referred to as certainty-equivalence regularization because it leads to the controller equivalent to the certainty-equivalence indirect data-driven LQR with least-square estimation of the system model when $\gamma$ is sufficiently large~\cite{Dorfler2022On}.
Meanwhile, another regularization that can guarantee robustness has been proposed:
\begin{equation}\label{eq:prob_reg2}
\begin{array}{cl}
 \displaystyle{\min_{P,K,G}} & \trace(QP)+\trace(K^{\sf T}RKP)+\rho\,\trace(GPG^{\sf T})\\
 {\rm s.t.} & X_1GPG^{\sf T}X_1^{\sf T}-P+I\preceq 0\\
  & P \succeq I\ {\rm and}\ \eqref{eq:KI}\\
\end{array}
\end{equation}
with a constant $\rho\geq0$.
The regularizer $\rho\,\trace(GPG^{\sf T})$ plays the role to reduce the size of the matrix $GPG^{\sf T}$ to achieve the actual stability requirement $(X_1-D_0)GPG^{\sf T}(X_1-D_0)^{\sf T}-P+I\preceq 0$ using the constraint $X_1GPG^{\sf T}X_1^{\sf T}-P+I\preceq 0$.
We refer to the latter one as robustness-inducing regularization.
For reformulation of~\eqref{eq:prob_ori},~\eqref{eq:prob_reg1}, and~\eqref{eq:prob_reg2} into convex programs, see~\cite{Dorfler2021Certainty}.

\subsection{Fast Gradient Sign Method}

The fast gradient sign method (FGSM) is a method to efficiently compute an adversarial perturbation for a given image~\cite{Goodfellow2015Explaining}.
Let $L(X,Y;\theta)$ be the loss function of the neural network where $X\in\mathcal{X}$ is the input image, $Y\in\mathcal{Y}$ is the label, and $\theta$ is the trained parameter, and let $f:\mathcal{X}\to\mathcal{Y}$ be the trained classification model. 
The objective of the adversary is to cause misclassification by adding a small perturbation $\Delta\in\mathcal{X}$ such that $f(X+\Delta)\neq f(X)$.
Specifically, the max norm of the perturbation is restricted, i.e., $\|\Delta\|_{\rm max} \leq \epsilon$
with a small constant $\epsilon>0$.

The core idea of FGSM is to choose a perturbation that locally maximizes the loss function.
The linear approximation of the loss function with respect to $\Delta$ is given by
\begin{equation}\label{eq:FGSM_linear_approx}
 L(X+\Delta,Y;\theta)\simeq L(X,Y;\theta)+\sum_{k,\ell}(\nabla_{X}L(X,Y;\theta))_{k\ell}\Delta_{k\ell}
\end{equation}
where the subscript $(\cdot)_{k\ell}$ denotes the $(k,\ell)$ component.
The right-hand side of~\eqref{eq:FGSM_linear_approx} is maximized by choosing $\Delta_{k\ell}=\epsilon\,{\rm sign}(\nabla_{X}L(X,Y;\theta))_{k\ell}$, whose matrix form is given by
\[
 \Delta = \epsilon\, {\rm sign}(\nabla_{X}L(X,Y;\theta)).
\]
FGSM creates a series of perturbations in the form increasing $\epsilon$ until misclassification occurs.
In the next section, we apply this idea to adversarial attacks on direct data-driven control for destabilization.

\section{ADVERSARIAL ATTACKS to DIRECT DATA-DRIVEN CONTROL}
\label{sec:attack}

\subsection{Threat Model}
\begin{figure}[t]
  \centering
  \includegraphics[width=0.98\linewidth]{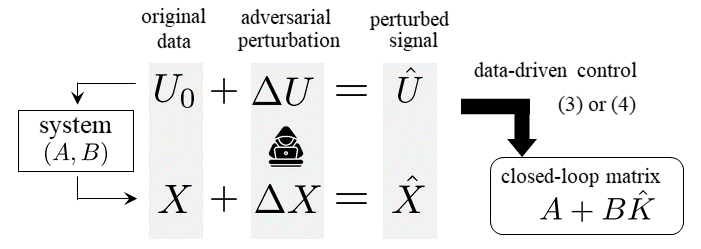}
  \caption{
  Threat model considered in this paper.
  The adversary is able to add a perturbation $(\Delta U, \Delta X)$ to the original input and output data $(U_0,X)$ with knowledge of the system model, the signals, and the controller design algorithm.
  The controller $\hat{K}$ is designed using the perturbed data $(\hat{U},\hat{X})$, which results in the closed-loop matrix $A+B\hat{K}$.
  }
  \label{fig:scenario}
\end{figure}

This study considers the following threat model:
The adversary can add a perturbation $(\Delta U, \Delta X)$ to the input and output data $(U_0,X)$.
Additionally, the adversary knows the system model $(A,B)$, the data $(U_0,D_0,X)$, and the controller design algorithm.
This scenario is depicted in Fig.~\ref{fig:scenario}.
The controller $\hat{K}$ is designed using the perturbed data $(\hat{U},\hat{X}):=(U_0+\Delta U,X+\Delta X)$, which results in the closed-loop matrix $A+B\hat{K}$.
The attack objective is to destabilize the system by crafting a small perturbation such that the closed-loop matrix has an eigenvalue outside the unit circle.

Additionally, we consider gray-box attacks where the adversary has access to the system model $(A,B)$ and the controller design algorithm but not the data $(U_0,D_0,X)$, and additionally may not know the design parameters $\gamma$ and $\rho$.
In this case, a reasonable attack strategy is to use hypothetical input $\hat{U}$ and disturbance $\hat{D}_0$ and calculate the corresponding state trajectory $\hat{X}$.
We refer to effectiveness of the attack without knowledge of the data as \emph{the transferability across data.}
Additionally, when the design parameters are unknown, hypothetical design parameters $\hat{\gamma}$ or $\hat{\rho}$ are also used.
We refer to the effectiveness in this scenario as \emph{the transferability across parameters.}
We numerically evaluate the transferability properties in Sec.~\ref{sec:ex}.

\subsection{Directed Gradient Sign Method}
We develop \emph{the directed gradient sign method (DGSM)} to design a severe perturbation
$\Delta:=(\Delta U, \Delta X)$ that satisfies $\|\Delta\|_{\rm max}\leq \epsilon$ with a small constant $\epsilon>0$.
Let
\[
 \Lambda(U_0,X,\Delta):=\sigma(A+B\hat{K})
\]
denote the eigenvalues of the closed-loop system with the direct data-driven control~\eqref{eq:prob_reg1} or~\eqref{eq:prob_reg2} using the perturbed data $(\hat{U},\hat{X})$.
The aim of the attack is to place some element of $\Lambda(U_0,X,\Delta)$ outside the unit circle.

\begin{figure}[t]
  \centering
  \includegraphics[width=0.98\linewidth]{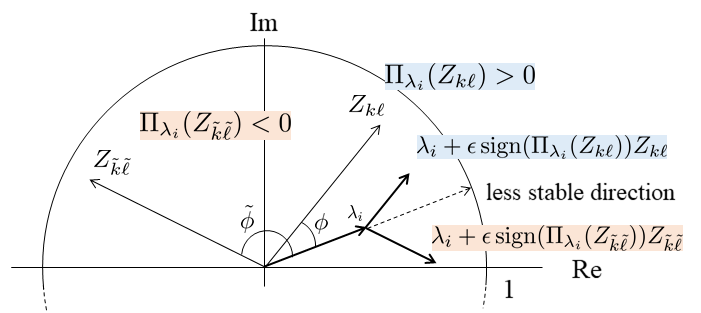}
  \caption{
  Role of the function $\Pi_{\lambda_i}$ in~\eqref{eq:Pi}.
  Since $Z_{k\ell}$ faces the direction of $\lambda_i$, the angle $\phi$ between $\lambda_i$ and $Z_{k\ell}$ is less than $\pi/2$, which leads to $\Pi_{\lambda_i}(Z_{k\ell})>0$.
  On the other hand, since $\tilde{\phi}$ between $\lambda_i$ and $Z_{\tilde{k}\tilde{\ell}}$ is greater than $\pi/2$, $\pi_{\lambda_i}(Z_{\tilde{k}\tilde{\ell}})<0$.
  As a result, in both cases, the perturbed eigenvalue moves closer to the unit circle.
  }
  \label{fig:Pi}
\end{figure}

The core idea of DGSM is to choose a perturbation that locally shifts an eigenvalue in the less stable direction.
We temporarily fix the eigenvalue of interest, denoted by $\lambda_i(U_0,X,\Delta)$,
and denote its gradient with respect to $\Delta$ by $\nabla_{\Delta}\lambda_i(U_0,X,\Delta)$.
The linear approximation of the eigenvalue with respect to $\Delta$ is given by
\begin{equation}\label{eq:DGSM_linear_approx}
 \lambda_i(U_0,X,\Delta)\simeq \lambda_i(U_0,X,0)+\sum_{k,\ell} \nabla_{\Delta}\lambda_i(U_0,X,\Delta)) \Delta_{k\ell}.
\end{equation}
We choose $\Delta_{k \ell}$ such that the right-hand side of~\eqref{eq:DGSM_linear_approx} moves closer to the unit circle.
Specifically, DGSM crafts the perturbation
\[
 \Delta = \epsilon\, {\rm sign}(\Pi_{\lambda_i}(\nabla_{\Delta}\lambda_i(U_0,X,\Delta)))
\]
where $\Pi_{\lambda_i}:\mathbb{C}^{(m+n)\times(2T+1)}\to \mathbb{R}^{(m+n)\times(2T+1)}$ is defined by
\begin{equation}\label{eq:Pi}
 \Pi_{\lambda_i}(Z):= \Real{\lambda_i}\Real{Z}+\Imag{\lambda_i}\Imag{Z}
\end{equation}
with
\[
 Z:=\nabla_{\Delta}\lambda_i(U_0,X,\Delta).
\]
The role of the function $\Pi_{\lambda_i}$ is illustrated in Fig.~\ref{fig:Pi}.
Suppose that $Z_{k\ell}$ faces the direction of $\lambda_i$.
More precisely, the angle between $\lambda_i$ and $Z_{k\ell}$, denoted by $\phi$, is less than $\pi/2$, which leads to $\Pi_{\lambda_i}(Z_{k\ell})>0$.
We now suppose that the angle between $\lambda_i$ and another element $Z_{\tilde{k}\tilde{\ell}}$, denoted by $\tilde{\phi}$, is greater than $\pi/2$.
Then we have $\Pi_{\lambda_i}(Z_{\tilde{k}\tilde{\ell}})<0$.
In both cases, owing to the function $\Pi_{\lambda_i},$ the perturbed eigenvalue moves closer to the unit circle as depicted in the figure.
By aggregating all components, the linear approximation of the perturbed eigenvalue $\hat{\lambda}_i$ is given  by
\[
 \textstyle{\hat{\lambda}_i\simeq\lambda_i+\epsilon\,\sum_{k,\ell}{\rm sign}(\Pi_{\lambda_i}(Z_{k\ell}))Z_{k\ell},}
\]
which is expected to be placed outside the unit circle by increasing $\epsilon$.

DGSM performs the procedure above for every $\lambda_i$ for $i=1,\ldots,n$ increasing $\epsilon$ until the resulting system is destabilized.
Its algorithm is summarized in Algorithm~1, where $\{\epsilon_k\}$ denotes possible candidates of the constant $\epsilon$ in the ascending order.
Algorithm~1 finds a perturbation $\Delta$ with the smallest $\epsilon$ in $\{\epsilon_k\}$ such that the resulting closed-loop system becomes unstable.

\begin{algorithm}[th]
\caption{Directed Gradient Sign Method (DGSM)}
\begin{algorithmic}[1]
\REQUIRE{$\{\epsilon_k\},A,B,U_0,X,\gamma,\rho$}
\ENSURE{$\Delta$}
\STATE ${\rm flag} \leftarrow 0$
\STATE $k \leftarrow 0$
\WHILE{${\rm flag} = 0$}
\STATE $k \leftarrow k+1$
\FOR{$i=1,\ldots,n$}
\STATE $\Delta \leftarrow \epsilon_k {\rm sign}(\Pi_{\lambda_i} (\nabla_\Delta \lambda_i(U_0,X,\Delta)) )$
\IF{$|\lambda_i(U_0,X,\Delta)|>1$}
\STATE ${\rm flag}\leftarrow 1$
\STATE \textbf{break}
\ENDIF
\ENDFOR
\ENDWHILE
\RETURN $\Delta$
\end{algorithmic}
\end{algorithm}

\section{NUMERICAL EXPERIMENTS}
\label{sec:ex}

\subsection{Experimental Setup}

We evaluate our adversarial attacks through numerical experiments.
We consider the inverted pendulum~\cite{Chalupa2008Modelling} with sampling period $0.01$ whose system matrices are given by
\[
  A=\left[
 \begin{array}{ccc}
 0.9844 & 0.0466 & 0.0347\\
 0.0397 & 1.0009 & 0.0007\\
 0.0004 & 0.0200 & 1.0000
 \end{array}
 \right],\quad B=\left[
 \begin{array}{c}
 0.25 \\
 0\\
 0
 \end{array}
 \right].
\]
We set the weight matrices to $Q=I$ and $R=10^{-5}I$.
The input signal is randomly and independently generated by $u(t)\sim \mathcal{N}(0,1)$.
We consider the disturbance-free case, i.e., $d(t)=0$.
The time horizon is set to $T=10$.
The $2$-induced norm is taken as the matrix norm in~\eqref{eq:prob_reg1}.
The gradient $\nabla_{\Delta}(\lambda_i(U_0,X,\Delta))$ is computed by the central difference approximation~\cite[Chapter~4]{Burden2015Numerical}.

\subsection{Robustness Improvement by Regularization}

We examine the improvement in robustness through regularization by comparing DGSM with a random attack where each element of $\Delta$ takes $\epsilon$ or $-\epsilon$ with equal probability.
Let $N_{\rm all}$ and $N_{\rm unstable}$ denote the total number of samples and the number of the samples where the resulting closed-loop system is unstable, respectively.
In addition, let $\bar{\epsilon}$ denote the minimum $\epsilon$ such that $N_{\rm unstable}/N_{\rm all}\geq\tau$ for a given threshold $\tau\in[0,1]$.
We set $N_{\rm all}=50$ and $\tau=0.8$.

\begin{figure}[t]
  \centering
  \includegraphics[width=0.98\linewidth]{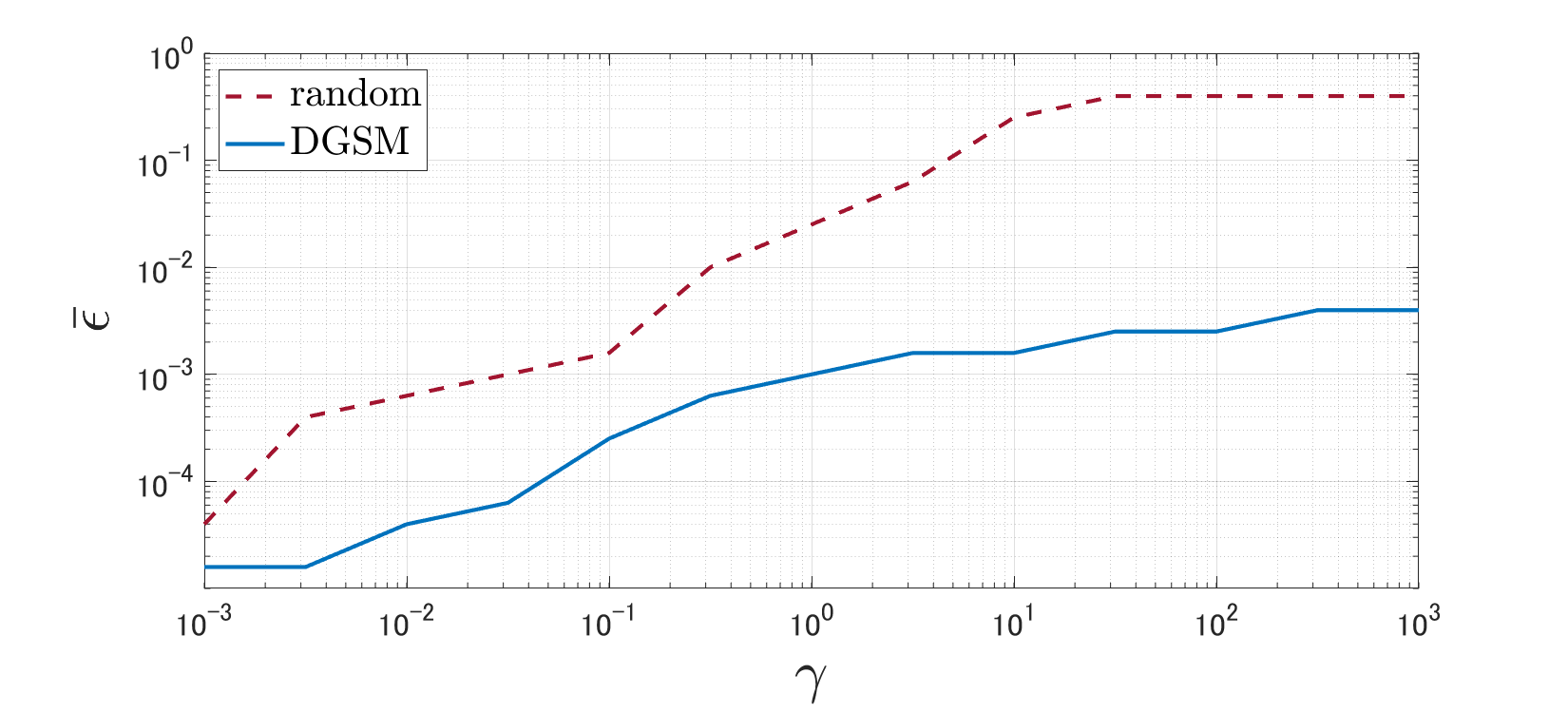}
  \caption{
  Curves of $\bar{\epsilon}$ with varying $\gamma$ for DGSM and the random attack when using the certainty-equivalence regularization~\eqref{eq:prob_reg1}.
  }
  \label{fig:eps_vs_lambda}
\end{figure}

Fig.~\ref{fig:eps_vs_lambda} depicts the curves of $\bar{\epsilon}$ with varying $\gamma$ for DGSM and the random attack when using the certainty-equivalence regularization~\eqref{eq:prob_reg1}.
First, it is observed that the magnitude of the adversarial perturbation necessary for destabilization increases as the regularization parameter $\gamma$ increases.
This result implies that the regularization method originally proposed for coping with disturbance is also effective in improving robustness against adversarial attacks.
Second, the necessary magnitude in DGSM is approximately $10\%$ of that in the random attack, which illustrates the significant impact of DGSM.

\begin{figure}[t]
  \centering
  \includegraphics[width=0.98\linewidth]{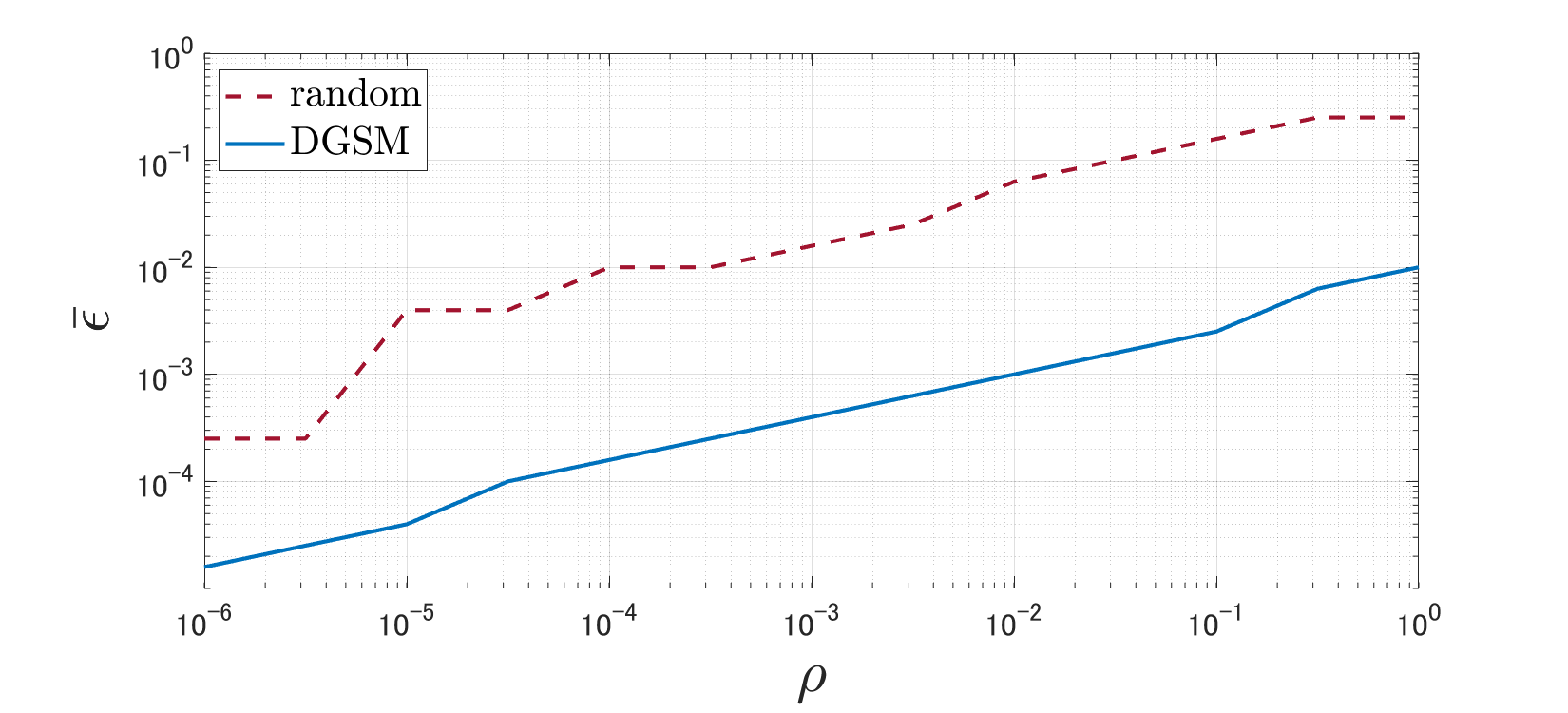}
  \caption{
  Curves of $\bar{\epsilon}$ with varying $\rho$ when using the robustness-inducing regularization~\eqref{eq:prob_reg2}.
  }
  \label{fig:eps_vs_rho}
\end{figure}

Fig.~\ref{fig:eps_vs_rho} depicts the curves of $\bar{\epsilon}$ with varying $\rho$ when using the robustness-inducing regularization~\eqref{eq:prob_reg2}.
This figure shows results similar to Fig.~\ref{fig:eps_vs_lambda}.
Consequently, both regularization methods are effective for adversarial attacks.

\begin{figure}[t]
  \centering
  \includegraphics[width=0.98\linewidth]{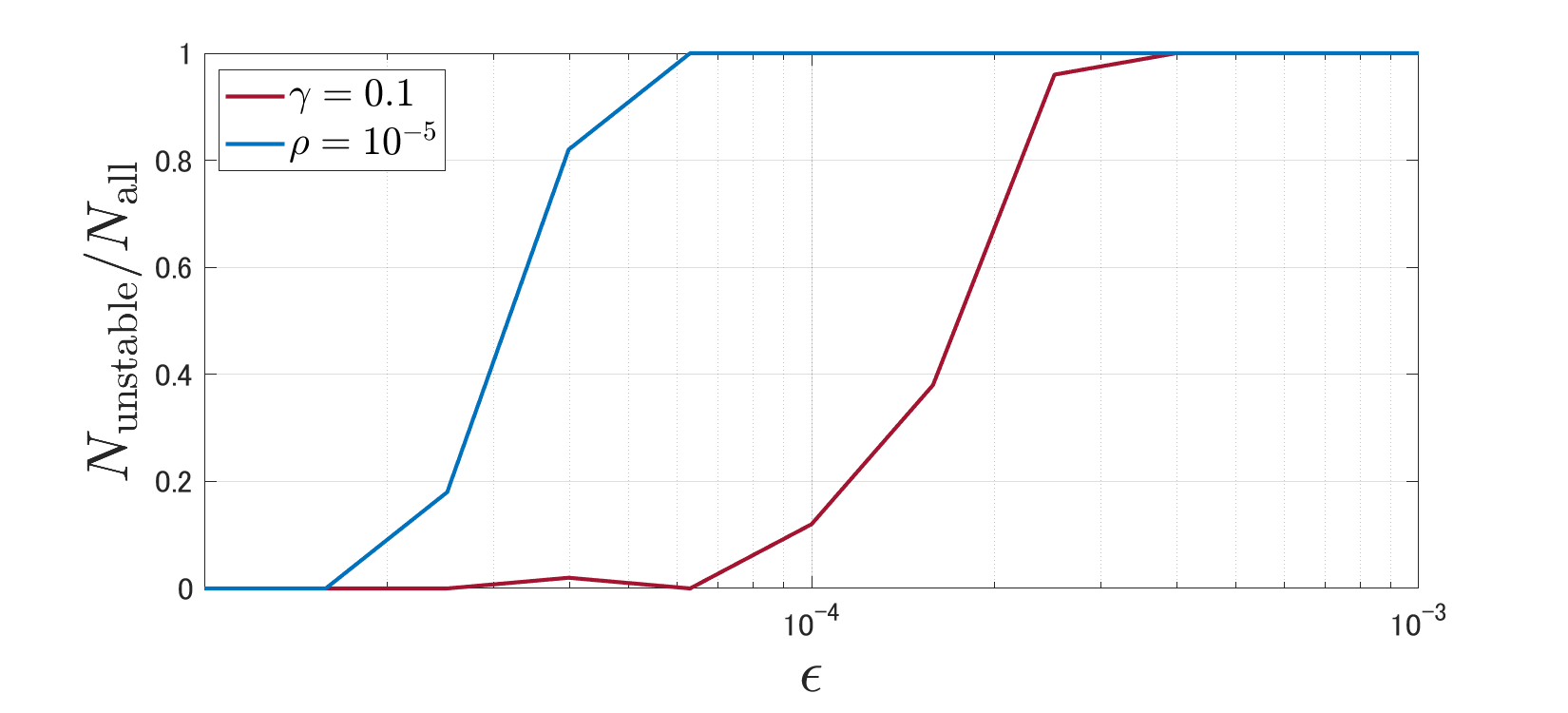}
  \caption{
  Ratio of unstable instances $N_{\rm unstable}/N_{\rm all}$ with the two regularized controller design methods~\eqref{eq:prob_reg1} and~\eqref{eq:prob_reg2} for varying $\epsilon$.
  }
  \label{fig:lam_rho_comp}
\end{figure}

Next, we compare the effectiveness of the two regularization methods.
We take $\gamma=0.1$ and $\rho=10^{-5}$ such that the resulting closed-loop performances $J(K)$ are almost equal.
Fig.~\ref{fig:lam_rho_comp} depicts $N_{\rm unstable}/N_{\rm all}$ with the two regularized controller design methods~\eqref{eq:prob_reg1} and~\eqref{eq:prob_reg2} for varying $\epsilon$.
It can be observed that the robustness-inducing regularization~\eqref{eq:prob_reg2} always outperforms the certainty-equivalence regularization~\eqref{eq:prob_reg1}.

\subsection{Transferability}

\begin{figure}[t]
  \centering
  \includegraphics[width=0.98\linewidth]{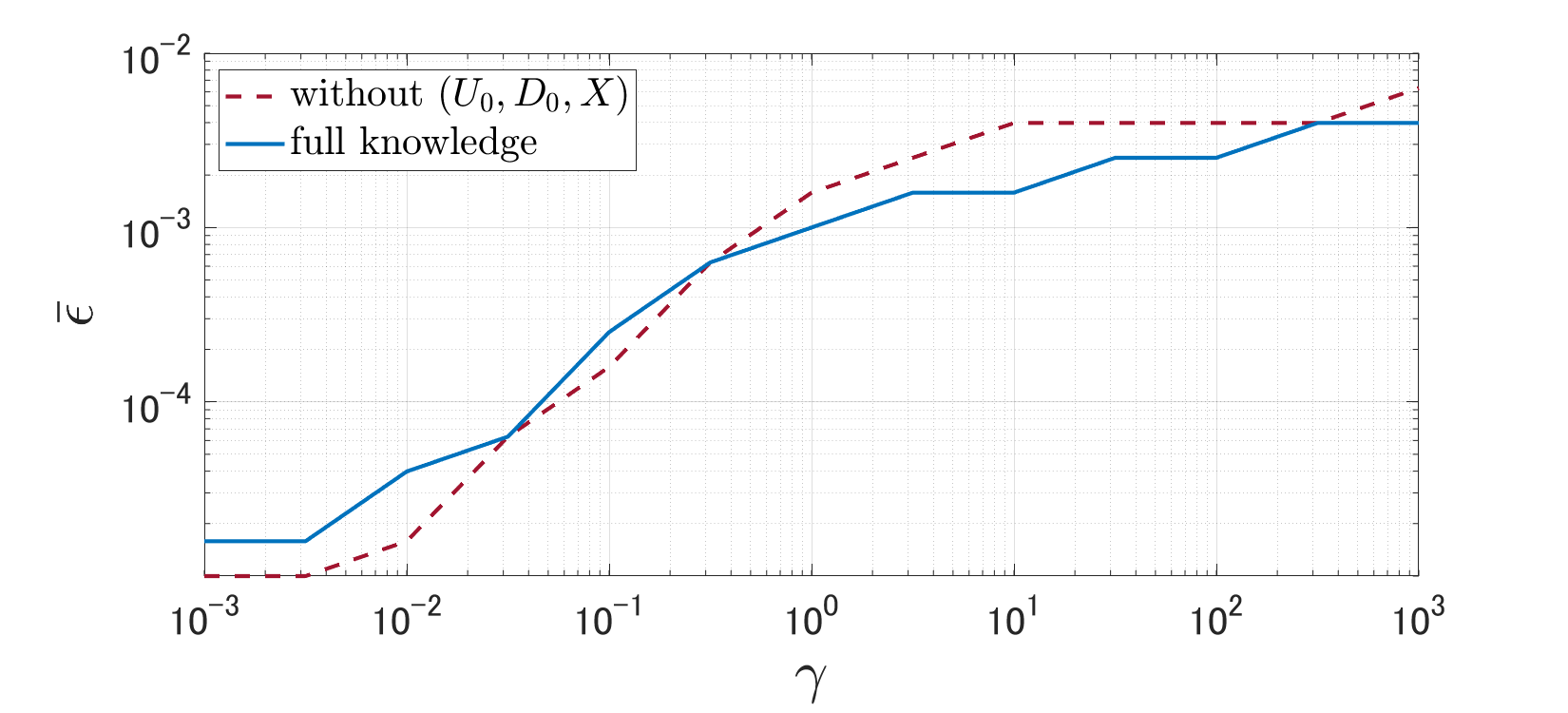}
  \caption{
  Curves of $\bar{\epsilon}$ with varying $\gamma$ for DGSM without knowledge of data and that with full knowledge.
  }
  \label{fig:trans_data}
\end{figure}

We consider transferability across data where the data $(U_0,D_0,X)$ is unknown and DGSM uses a hypothetical input $\hat{U}_0$ whose elements are also randomly and independently generated by $\mathcal{N}(0,1)$ and $\hat{D}_0=0$.
Fig.~\ref{fig:trans_data} depicts the curves of $\bar{\epsilon}$ with varying $\gamma$ for DGSM without knowledge of data and that with full knowledge when using the certainty-equivalence regularization~\eqref{eq:prob_reg1}.
This figure shows that DGSM exhibits the transferability property across data.

\begin{figure}[t]
  \centering
  \includegraphics[width=0.98\linewidth]{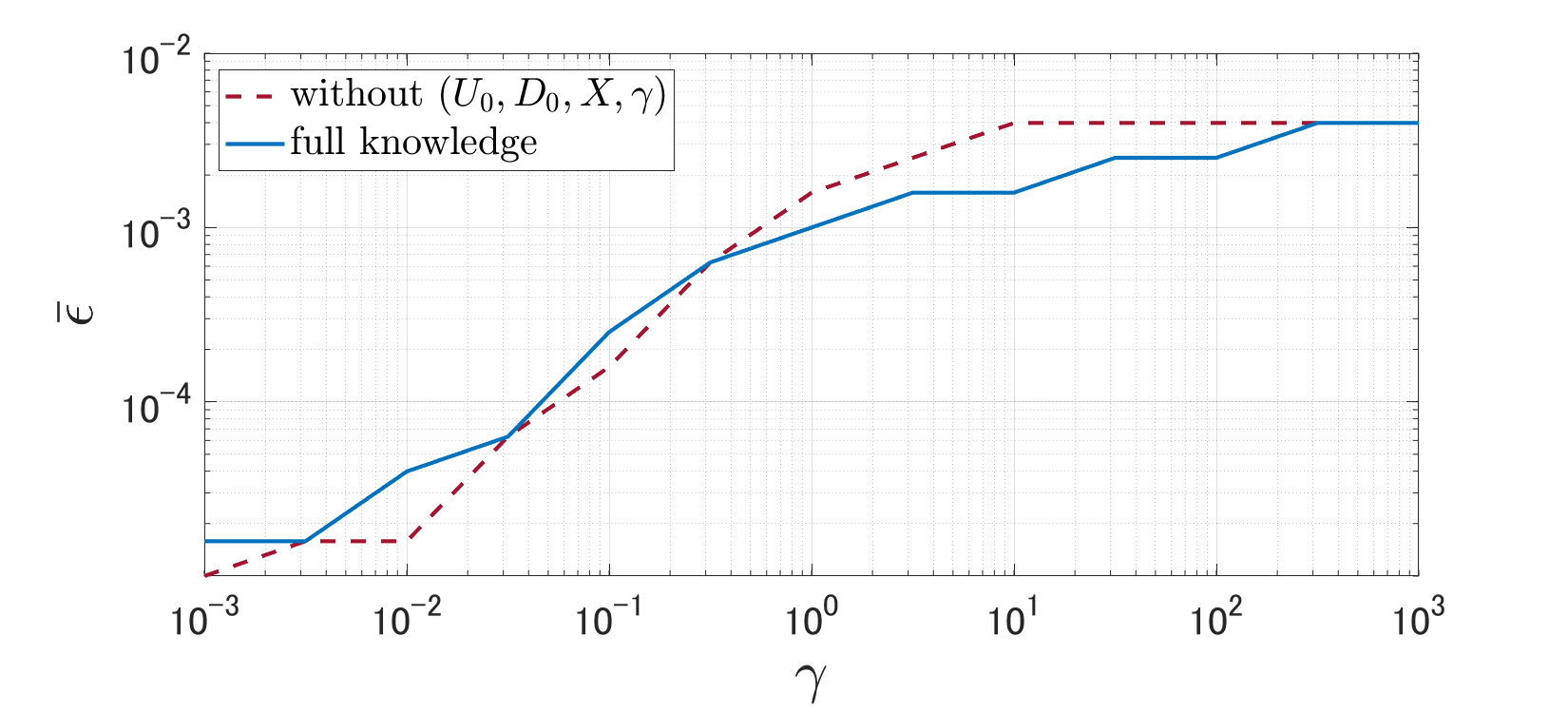}
  \caption{
  Curves of $\bar{\epsilon}$ with varying $\gamma$ for DGSM without knowledge of data and parameters and that with full knowledge.
  }
  \label{fig:trans_para}
\end{figure}

Subsequently, we examine transferability across design parameters where the regularization parameter $\gamma$ in addition to the data $(U_0,D_0,X)$ is unknown.
We use $\gamma=0.1$ as a hypothetical parameter.
Fig.~\ref{fig:trans_para} depicts the corresponding curves
As in the transferability across data, the results confirm the transferability property across parameters.

\subsection{Discussion}

The regularization methods described by~\eqref{eq:prob_reg1} and~\eqref{eq:prob_reg2} provide a quantitative condition to ensure stability:
The resulting closed-loop system with the certainty-equivalence regularization is stable when $\gamma$ and the signal-to-noise ratio (SNR) defined by
${\rm SNR}:=\sigma_{\rm min}(W_0)/\sigma_{\max}(D_0)$
are sufficiently large~\cite[Theorem~4.2]{Dorfler2021Certainty}.
That with the robustness-inducing regularization is stable when $\rho$ is sufficiently large and $\sigma_{\rm max}(D_0)$ is sufficiently small~\cite[Theorem~3]{De2021Low}.
One may expect that DGSM crafts a severe input perturbation such that its maximum singular value is large but its elements are small.
However, for the single-input system, $\sigma_{\rm max}(\Delta U)=\epsilon \sqrt{T}$ for any $\Delta U$ whose elements take $\epsilon$ or $-\epsilon$.
This means that the input perturbations made by DGSM and the random attack have the same maximum singular value.

\section{CONCLUSION}
\label{sec:conc}

This study has investigated the vulnerability of direct data-driven control, specifically focusing on the Willems' fundamental lemma-based approach with two regularization methods, namely certainty-equivalence regularization and robustness-inducing regularization.
To this end, a new method called DGSM, based on FGSM which has been originally been proposed for neural networks, has been introduced.
It has been demonstrated that direct data-driven control can be vulnerable, i.e., the resulting closed-loop system can be destabilized by a small but sophisticated perturbation.
Numerical experiments have indicated that strengthening regularization enhances robustness against adversarial attacks.

Future research should include further tests of the vulnerability with various types of data and systems under different operating conditions, a theoretical analysis of DGSM, and exploration of novel defense techniques for reliable direct data-driven control.
For example, detection of adversarial perturbations~\cite{Metzen2017Detecting} is a promising direction.
Finally, for a more comprehensive understanding of the vulnerability, more sophisticated attacks should be considered.

\appendix
The parameters in the simulation in Fig.~\ref{fig:vuln_ins} are as follows.
The system is a marginally unstable Laplacian system
considered in~\cite{Dorfler2022On,Dean2020Sample}.
Each element of the disturbance $D$ is randomly generated from $\mathcal{N}(0,d^2)$ with $d=0.05$.
The weight matrices are $Q=I$ and $R=10^{-3}I$.
The time horizon is set to $T=15$.
The magnitude of the adversarial perturbation is set to $\epsilon=0.16$.
The controller is designed using the certainty-equivalence regularization~\eqref{eq:prob_reg1} with a regularization parameter $\gamma=10^{-3}$.

\bibliographystyle{IEEEtran}
\bibliography{sshrrefs}

\end{document}